\documentclass[onecolumn]{mn2e}
\usepackage{graphicx}
\usepackage[section]{placeins}

\pagerange{\pageref{firstpage}--\pageref{lastpage}} \pubyear{2009}

\begin{document}

\title[On the aberration/retardation effects in radio pulsars]{On the
aberration--retardation effects in pulsars.}
\author[Krzeszowski et. al.]{K.~Krzeszowski,$^1$ D.~Mitra,$^2$ Y.~Gupta,$^2$
J.~Kijak,$^1$ J.~Gil,$^1$ and A.~Acharyya$^3$ \\
$^1$J. Kepler Institute of Astronomy, University of Zielona G\'ora, Lubuska 2, 65--265,
Zielona G\'ora, Poland\\
$^2$National Centre for Radio Astrophysics, TIFR, Pune University Campus,
Post Bag 3, Ganeshkind PO, Pune Maharahtra 411007, India\\
$^3$School of Electronics and Computer Science, University of Southampton,
Southampton -- SO17 1BJ, UK\\
}
\date{Released 2009}
\maketitle

\begin{abstract}
The magnetospheric locations of pulsar radio emission region are not
well known. The actual form of the so--called radius--to--frequency
mapping should be reflected in the aberration--retardation (A/R)
effects that shift and/or delay the photons depending on the
emission height in the magnetosphere. 
Recent studies suggest that in a handful of pulsars
the A/R effect can be discerned w.r.t the peak of the 
central core emission region.
To verify these effects in an
ensemble of pulsars we launched a project analysing multi--frequency
total intensity pulsar profiles obtained from the new observations
from the Giant Meterwave Radio Telescope (GMRT), Arecibo Observatory
(AO) and archival European Pulsar Network (EPN) data. For all these
profiles we measure the shift of the outer cone components with
respect to the core component which is necessary for establishing
the A/R effect. Within our sample of 23 pulsars 7 show the A/R
effects, 12 of them (doubtful cases) show a tendency towards
this effect, while the remaining 4 are obvious counter examples. The 
counter--examples and doubtful cases may
arise from uncertainties in determination of the location of the
meridional plane and/or the core emission component. It hence appears that
the A/R effects are likely to operate in most pulsars from our
sample. We conclude that in cases where those effects are present
the core emission has to originate below the conal emission region.
\end{abstract}

\label{firstpage}


\section{Introduction}

Aberration and retardation effects (A/R hereafter) can be observed in a pulsar profile as a
shift of the position of conal components with respect to the core component
towards earlier longitudinal phases (see for e.g. Malov \& Suleimanova (1998), Gangadhara \& Gupta 2001,
G\&Ga hereafter).
Such effects should occur if different
components are emitted at different distances from the pulsar surface
(emission radii), as well as from the pulsar spin axis. Aberration is caused
by bending of radiation beam due to the polar cap rotation, while
retardation is based on a path difference for radiation from different conal
emission regions to reach an observer. If we assume that the emission from
the core component arises relatively close to the star surface, then it
should not be strongly affected by either of the two above mentioned
effects. This will be our initial assumption.

To determine A/R shifts the pulsar profile has to meet certain requirements.
It has to be a high resolution profile with high signal to noise (S/N) ratio.
The core and the conal components have to be clearly identified within the
profile. Multifrequency data is recommended, so one can follow
the profile evolution throughout all frequencies, which can help to identify
different emission components.

When values of A/R shifts are determined, then the heights of the radiation
emission region (emission altitudes hereafter) can be calculated (see
G\&Ga and Dyks et. al 2004). It is
believed that at different frequencies the emission arises at different
heights above the pulsar surface (Kijak \& Gil 1998, 2003 and Mitra \&
Rankin 2002). The results of this analysis can be used to verify the existence of
a radius--to--frequency mapping. All observational limits for emission
altitude hence can be crucial for understanding the physical mechanism of
generation of pulsar coherent radio emission.

The relativistic beaming model initially proposed by Blaskiewicz, Cordes \&
Wasserman (1991, BCW hereafter) clearly demonstrated that aberration and
retardation effects play an important role in pulsars. This study was
primarily based on evidence which followed from the effects of A/R as seen
in the inflexion point of the pulsar's polarisation position angle (hereafter PA) traverse, which lags
the midpoint of the total intensity profile centre. A similar effect of A/R
was reported by G\&Ga and Gupta \& Gangadhara (2003, G\&Gb hereafter) in a
handful of pulsars where the core emission was seen to lag behind the
profile centre.

In this paper we have undertaken a careful study to establish the A/R effect
observed by G\&Ga and G\&Gb for a large sample of pulsars observed at multiple
frequencies. Most of the data are new observations from the Giant Meterwave
Radio Telescope (GMRT hereafter) and the Arecibo Observatory (AO hereafter).
We have also used some archival data from the European Pulsar Network (EPN
hereafter) archive\footnote{%
http://www.mpifr--bonn.mpg.de/div/pulsar/data/}. In section~(\ref{sec2}) we
discuss various methods used to find emission heights in pulsars, in section~(\ref{sec3})
we discuss various factors affecting A/R measurements in pulsars
and section~(\ref{sec4}) deals with the observation and data analysis methods used
in this paper.

As a result of our analysis presented in section~(\ref{sec5}) we found that
out of 23 pulsars in our sample 7 clearly show the A/R effect, 12 show a
clear tendency towards this effect, while the remaining 4 are counter
examples. However, as argued in section~(\ref{sec3}), all problematic cases
(pulsar profiles at all or some frequencies not showing the A/R effect) can
be attributed to a number of effects like incorrect identification of the core
component or missing conal emission. We can conclude that A/R effects are seen to operate in
pulsars, which we discuss in section~(\ref{sec6}).

\section{Geometry and emission altitudes}
\label{sec2}
Radio emission heights in pulsars are primarily obtained by two methods; the geometrical
method and heights estimation based on A/R effects.
Here we briefly mention the essential ingredients of the methods used,
and a detailed discussion of the various
methods used can be found in Mitra \& Li (2004).

 Radio emission geometry is determined by several parameters: $%
\alpha$ --- an inclination angle of the magnetic dipole with respect to the
rotation axis, $\beta$ --- the minimum angle between an observer's line of
sight and magnetic axis (impact angle), $\rho$ --- an opening angle of the
radio emission beam, $r$ --- a radial distance of the radio emission region
measured from the centre of the neutron star (emission altitude). The
opening angle $\rho$ of the pulsar beam corresponding to the pulse width $W$
is given by:
\begin{equation}  \label{rho_1}
\rho = 2 \arcsin\left(\sqrt{\sin\left(\alpha +
\beta\right)\cdot\sin\alpha\cdot\sin^2\frac{W}{4}+\sin^2\frac{\beta}{2}}%
~~\right)~.
\end{equation}
where $\alpha$, $\beta$, $W$ and $\rho$ are measured in degrees (Gil et al. 1984).
The opening angle $\rho$ is the angle between the pulsar
magnetic axis and the tangent to magnetic field lines at points where the
emission corresponding to the apparent pulse width $W$ originates. For
dipolar field lines:
\begin{equation}  \label{rho}
\rho = 1.24^\circ s (r/10^6 \mathrm{cm})^{1/2}P^{-1/2}~,
\end{equation}
(Gil \& Kijak 1993), where $s = d/r_\mathrm{p}$ is a mapping parameter which
describes the locus of corresponding field lines on the polar cap ($s = 0$
at the pole and $s = 1$ at the edge of the polar cap), $d$ is the distance
of the given magnetic open field line from the dipolar magnetic axis (in cm), $r_%
\mathrm{p}$ is the polar cap radius (in cm) and $P$ is the pulsar period in seconds.
The radio emission altitude can be obtained using eqn.~(%
\ref{rho}):

\begin{equation}  \label{r_geo}
r_\mathrm{geo} \simeq 10~ P \left(\frac{\rho}{1.24^\circ} \right)^2~\mathrm{km}.
\end{equation}
In this equation parameter $s = 1$ is used which corresponds to the
bundle of last open magnetic field lines. Kijak and Gil (1997) also derived a
semi--empirical formula for emission height which was slightly modified by
Kijak \& Gil (2003) by using larger number of pulsars and broadening the
frequency coverage in their analysis. They estimated the emission heights for a
number of pulsars at several frequencies using the geometrical method
(eqs. \ref{rho_1} and \ref{r_geo}) based on the above assumption that the low intensity pulsar radiation
(corresponding to the profile wings) is emitted tangentially to
the bundle of the last open dipolar field lines. In the method, they used
low--intensity pulse width data at about 0.01 \% of the maximum intensity
(Kijak \& Gil 1997 and Kijak \& Gil 2003). The most recent formula has a form
\begin{equation}  \label{rkg}
r_\mathrm{KG} = (400 \pm 80 )\nu^{-0.26 \pm 0.09}_{\mathrm{GHz}%
} \dot{P}^{0.07 \pm 0.03}_{\mathrm{-15}} P^{0.30 \pm 0.05}~\mathrm{km},
\end{equation}
where $\nu_{GHz}$ is observing frequency in GHz and $\dot{P}_{-15} = \dot{P}%
/10^{-15}$ is the period derivative. The formula $r_\mathrm{KG}$ sets lower and
upper boundaries for emission height values in the statistical sense.

For calculating the emission heights one can take a different approach.
The A/R effects have been observed in the total intensity profiles as discussed
by Malov \& Suleimanova (1998), G\&Ga, G\&Gb, Dyks et al (2004). In turn, they have been used to
calculate radio emission heights. Observationally the A/R effects are obtained by
measuring the positions of the conal emission components and identifying the central
core emission component.
The shape of the observed profile depends on the structure of radio beam, which
is thought to be composed of a central core emission surrounded by nested
cones (Rankin 1983, Mitra \& Deshpande 1999). The line of sight passing through
the central region of this beam will hence cut the cone emission at the edges
including the central core region. Following this, pulsar profiles can be classified due to
the number and type of visible components (Rankin 1983). If we assume that in
a certain profile the core component and both the conal components are
present, we can define the phase difference:
\begin{equation}  \label{delta_phi}
\Delta \phi = \frac{\phi_t - |\phi_l|}{2}~,
\end{equation}
where $\phi_l$ and $\phi_t$ are phase positions (in degrees) of the peaks of the leading and trailing
components with respect to the core phase ($\phi_c = 0^\circ$). Following
the refinement of G\&Ga by Dyks et al. (2004) one can estimate the radio
emission altitude\footnote{%
Recently in a detailed study Gangadhara (2005) has obtained a revised
expression (their eqn. 34) for the relativistic phase shift which depends on
the angles $\alpha$ and $\beta$, and the corrections in obtaining emission
heights turns out to be of the order of 2\% to 10\% for normal pulsars. We
have however not used this formula since all the pulsars in our sample are
normal pulsars and, further, the error in estimating emission heights are
often more than 10\%.}:
\begin{equation}  \label{r_del}
r_\mathrm{AR} \simeq -\frac{\Delta\phi}{2} \frac{2\pi} {360^\circ} R_{\mathrm{LC}}~,
\end{equation}
where $R_{\mathrm{LC}} = cP/2\pi$ (where $c$ is speed of light) is the light
cylinder radius. The A/R method has an edge over the geometrical method since it
only depends weakly on the geometrical angles of the star which are difficult
to find. Generally the emission altitudes found using both these methods are of
the order of few hundred kilometers above the surface of the neutron star.
In this paper we use equation (\ref{r_del}) for estimating emission
altitudes following the method outlined by G\&Ga. The various factors affecting usage of this
method are discussed in section (\ref{sec3}).

\section{Factors Affecting A/R measurements in pulsars}
\label{sec3}

\subsection{Location of the meridional point}
The crucial aspect for estimating radio emission heights using the A/R technique is to
identify the location of the meridional plane i.e. the plane containing the magnetic
axis and the pulsar rotation axis.

Generally polarisation observations have been used
to identify the meridional plane as per the rotating vector model (RVM) proposed by
Radhakrishnan \& Cooke (1969). According to the RVM the variation of the PA of the linear
polarisation across the pulse reflects  the underlying structure of the dipolar magnetic field
if one is located on the pulsar frame. In this case
the steepest gradient or the inflexion point of the PA traverse is
the point contained in the meridional plane, and should be coincident
with the midpoint of the pulse profile. However in the observers frame, as was
pointed out by BCW, Dyks et al (2004), the A/R effects causes the PA inflexion point
to appear delayed w.r.t. the midpoint of the pulse profile. The magnitude of the delay
is proportional to the height of emission from the neutron star surface and inversely
proportional to the pulsar period. Hence in the observer's plane the inflexion point
of the PA traverse is no longer the point containing the meridional plane, rather
the meridional plane lies somewhere in between the profile midpoint and the
inflexion point of the PA traverse.

\begin{figure}
\begin{center}
\includegraphics[height=6.5cm,angle=270]{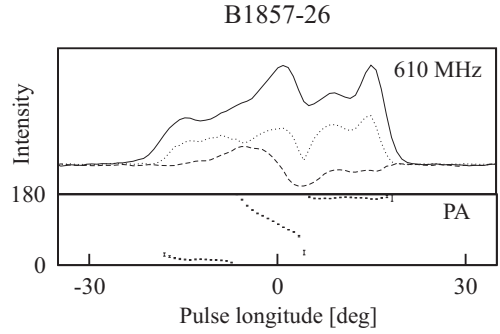}
\end{center}
\caption{B1857$-$26 polarisation profile. The upper panel of the plot shows the intensity (solid line), linear polarisation
(dotted line) and the circular polarisation (dashed line). The bottom panel shows the PA with error bars.
This profile is an example where the core component, the inflection point and the sign changing circular are close
to each other (see section 3 for details).}
\label{b1857}
\end{figure}

We use the method employed by G\&Ga,b to find emission heights due to
A/R effects using eq. (6).  If the mid--point of the outer conal peaks leads the core peak, it would
imply that the conal emission is higher than the core emission and the shift
can be used to find emission heights w.r.t. the core. This argument however assumes
that the meridional plane passes through the peak of the core emission.
While this assumption is not rigorously verified, there are few pulsars which have been used to
justify this argument. One such pulsar is PSR B1857$-$26 (see Fig. \ref{b1857}) where the peak of the core,
the inflexion point and the sign changing circular, are all close to each other (see Mitra
\& Rankin 2008 for a detailed discussion). If one assumes that the core emission is curvature
radiation which fills the full polar
cap as was suggested by Rankin (1990), then the meridional plane would lie
in the plane containing the inflexion point, peak of the core and sign changing
circular. Although a critical look at PSR B1857$-$26 reveals that the peak of the
core is delayed by a degree or so w.r.t. the inflexion point (see Mitra \& Rankin 2008).
Mitra \& Li (2004) have shown that  for a few other pulsars the peak of the core lags the
inflexion point by about a degree.
The shifts expected due to A/R effects for slow pulsars are also often small (e.g.
for emission altitude of 300 km and a 0.7 sec pulsar, the expected shift is
of the order of a degree) and is comparable to core shifts seen w.r.t. the inflexion point.
Hence, it is important to note that to demonstrate the A/R effects as suggested by G\&Ga and G\&Gb,
the meridional plane should pass through the peak of the core emission (which is
currently not verified) and the
conal emission should be rising higher than the core emission.

\subsection{Identifying the Core Emission}
Identification of the core emission is not trivial in pulsars.
Under the core--cone model of pulsar beaming the core is centrally located and surrounded by nested cones.
This implies that the core is located in the centre of a pulse profile, which is
very close to the inflexion point of the PA. Since the impact parameter
 needs to be small to cut through the core emission region in a pulsar,
the corresponding PA traverse as per the RVM should show a significant rotation
across the pulse. Core emission is often associated with a sign changing
circular polarisation and has a steeper spectral index than the conal emission.

\begin{figure}
\begin{center}
\includegraphics[height=6.5cm,angle=270]{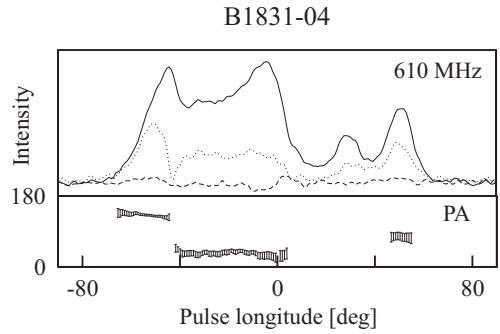}
\end{center}
\caption{B1831$-$04 polarisation profile (same as in Fig~\ref{b1857}). Although this pulsar shows the sign
change in circular it has a flat PA and a wide profile. Hence it is likely that the central bright emission component
is not a core (see section 3 for details)}
\label{b1831}
\end{figure}

There are situations where several core like emission characteristics are seen in a pulse component.
For example in Fig (\ref{b1831}) PSR B1831$-$04 has a bright central component which is
identified as core emission by Rankin (1993a, 1993b). This
component has a steep spectral index and a sign changing circular as seen in Gould \& Lyne's (1998, GL98 hereafter)
multi--frequency polarisation profiles. On the other hand the PA traverse is flat, showing no significant rotation
(except 90$^{\circ}$ jumps) and the profile is extremely wide ($\sim$ 150$^{\circ}$).
Such emission properties are consistent with the observer crossing the pulsar emission
beam along the conal emission zone and the observer is likely to miss the core emission.
Hence for PSR B1831$-$04, the bright centrally located component can be interpreted as
conal emission rather than core emission.
Further there are situations where the PA below the core emission
can deviate from the RVM model (see Mitra, Rankin \& Gupta 2007), and hence the location of
the inflexion point is difficult to determine.
A careful examination using multi--frequency single pulse polarisation data is hence essential to find
the location of the core emission.
For our data set we have used the GL98 polarisation data sets
wherever possible to find the location of the core emission which we discuss in the appendix.
However incorrect identification of the core would clearly affect demonstration of the A/R effect.

In several cases the central emission close to the core is weaker compared to the outer cones.
Methods employed to find the peak of the core (like the Gaussian fitting scheme discussed in section
\ref{sec4}) have problems in establishing the location of the core component (like PSR B1738$-$08), and
hence cause complication in discerning the A/R effect.

Subpulse modulation (or drifting) provides another means to identify core and conal components in pulsars. 
The conal components tends to show subpulse drifting seen as a spectral feature in phase resolved 
fluctuation spectra, while in the core emission this phenomenon is generally absent (Rankin 1986). 
It is sometimes possible that the 
central component of the pulsar shows drifting properties, in which case the central component 
cannot be core emission.  We have verified our sample pulsars for such 
drifting of the central component based on the dataset of pulsar subpulse modulation 
published by Weltevrede et al. (2006, 2007).
Three pulsars from our sample show central component modulation feature out of which
two pulsars PSR B1508+55 and B1917+00 has a broad modulation feature which is consistent with core 
emission, while PSR B1944+17 shows clear drifting properties making this pulsar unsuitable for A/R analysis.

\subsection{Partial cone emission}

\begin{figure}
\begin{center}
\includegraphics[height=6.5cm,angle=270]{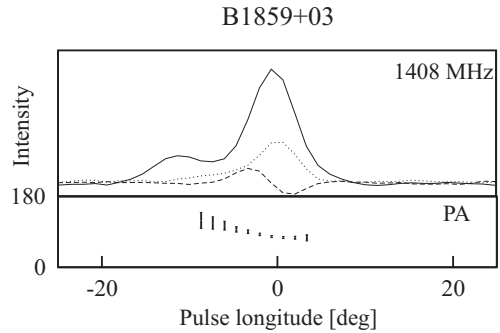}
\end{center}
\caption{B1859+03 polarisation profile (same as in Fig~\ref{b1857}). This pulsar has a clear sign changing circular
indicating the location of the core emission. However the emission towards the trailing parts of the profile seems to
be missing (see section 3 for details).}
\label{b1859}
\end{figure}

In several pulsars only a part of the emission cone is visible and these pulsars
were identified as partial cones by Lyne \& Manchester (1988). The PA inflexion point in these
pulsars lies towards one edge of the profile. This kind of pulsar however poses a problem
for finding the A/R effect, since estimation of the midpoint of the total intensity profile
is incorrect due to missing emission. One such example in our sample is perhaps PSR B1859+03
as shown in Fig (\ref{b1859}). The PA inflexion point coincides with the peak of the trailing component and
has a sign changing circular, which is suggestive of core emission, however there is no emission
component seen in the trailing part of the profile. Single pulse study by
Mitra, Rankin \& Sarala (2008) recently
has shown that the missing partial cone regions occasionally flare up, and these flared
profiles can be used to demonstrate the A/R effect.

\section{Observations and data analysis}

\label{sec4} For our analysis we selected profiles of 28 pulsars from three
sources (see Table \ref{tabelaPulsarow}): GMRT, AO and EPN. The GMRT (Swarup
et al. 1991) operating at meter wavelengths is an array of 30 antennas with
alt--az mounts, each of 45 meter diameter, spread out over a 25 km region
located 80 km north of Pune, India. It is designed to operate at multiple
frequencies (150, 235, 325, 610 and 1000 --- 1450 MHz respectively) having a
maximum bandwidth of 32 MHz split into upper and lower sidebands of 16 MHz
each. It is primarily an aperture synthesis interferometer as well as a
phased array instrument. The GMRT observations reported in
this paper are done in the phased array mode of operation at RF 225 and 325 MHz,
respectively, using the upper side band and a bandwidth of 16 MHz. In the phased array
mode of the GMRT (Sirothia, S. 2000, Gupta et al. 2000), signals from any
selected set of the antennas are added together after an appropriate delay and
phase compensation to synthesise a single, larger antenna with a narrower
beam. Signals of 16 MHz from each side band of each antenna are available
across 256 channels which can be summed in the GMRT array combiner. The
summed voltage signal output from the phased array is then processed like any other
single dish telescope signal, by the pulsar backend. At the frequencies of
interest here, GMRT has two linearly polarised outputs for each antenna
which are further mixed in a quadrature hybrid to produce two circular
polarised signals which go further into the front end electronics and rest
of the signal chain. The raw signals are finally post integrated by the
pulsar backend and the data was recorded with the final sampling time of
0.512 msec. The time resolution for all the analysis done in this paper
is the pulsar period divided by the sampling time.
All the pulsars were observed for about 2000 pulse periods to
obtain stable pulse profiles. The off--line analysis involved interference
rejection, dedispersion and folding to obtain the average pulse profiles.

The AO data set is from observations at the Arecibo Observatory, carried out
between December 2001 and June 2002. The pulsars were observed at different
frequencies using the following receivers : 430 MHz, L--band (1175 MHz),
S--low (2250 MHz), S--high (3500 MHz), C--band (4850--5000 MHz). The pulsar
back--end used most frequently was the WAPP (Wideband Arecibo Pulsar Processor), 
though the PSPM (Penn State Pulsar Machine) was also used
for some of the observations. The bandwidths of the observations were
typically 25/50 MHz at 430 MHz and 100 MHz for all the higher frequency
observations. The time resolution of the WAPP raw data was typically 128,
256 or 512 microsec. The data were analysed using the SIGPROC analysis
package to get the average folded profiles.

The data from the EPN archive were mostly obtained by GL98
using 76--m Lovell telescope at Jodrell Bank Observatory, except B1039$-$19
at 4850 MHz, which was obtained by Kijak et al. (1997) using Effelsberg
Observatory. Generally, resolution of EPN data is around 1$^\circ$, which is
much lower than that of GMRT or AO data.

\begin{table*}
\caption{List of pulsars chosen for analysis. The AO and GMRT observations are new,
whereas the EPN profiles have been taken from the data archive http://www.mpifr-bonn.mpg.de/div/pulsar/data/.}
\label{tabelaPulsarow}
\begin{center}
\begin{tabular}{c|c|c||c|c|c||c|c|c}
\hline
pulsar & $\nu$ (MHz) & data source & pulsar & $\nu$ (MHz) & data source &
pulsar & $\nu$ (MHz) & data source \\ \hline
B0402+61 & 225 & GMRT & B1541+09 & 225 & GMRT & B1911+13 & 325 & GMRT \\
& 408 & EPN &  & 325 & GMRT &  & 445 & AO \\
B0410+69 & 225 & GMRT &  & 457 & AO &  & 1225 & AO \\
& 408 & EPN &  & 1225 & AO & B1916+14 & 445 & AO \\
B0621$-$04 & 325 & GMRT &  & 2300 & AO &  & 1225 & AO \\
& 410 & EPN &  & 4900 & AO & B1917+00 & 225 & GMRT \\
& 610 & EPN & B1737+13 & 225 & GMRT &  & 325 & GMRT \\
& 1408 & EPN &  & 445 & AO & B1944+17 & 225 & GMRT \\
B0656+14 & 445 & AO &  & 457 & AO &  & 325 & GMRT \\
B0950+08 & 455 & AO &  & 1225 & AO & B1946+35 & 2300 & AO \\
B1039$-$19 & 225 & GMRT & B1738$-$08 & 225 & GMRT & B2002+31 & 1225 & AO \\
& 325 & GMRT &  & 325 & GMRT &  & 2300 & AO \\
& 1408 & EPN & B1821+05 & 225 & GMRT & B2020+28 & 445 & AO \\
& 4850 & EPN &  & 445 & AO & B2053+21 & 445 & AO \\
B1237+25 & 157 & GMRT &  & 1225 & AO & B2053+36 & 1225 & AO \\
& 455 & AO &  & 1460 & AO & B2315+21 & 225 & GMRT \\
& 1225 & AO &  & 2200 & AO &  & 1225 & AO \\
& 2300 & AO & B1831$-$04 & 225 & GMRT & B2319+60 & 225 & GMRT \\
& 3550 & AO &  & 325 & GMRT &  & 610 & EPN \\
& 4900 & AO &  & 606 & EPN &  & 925 & EPN \\
& 5050 & AO & B1839+09 & 2200 & AO &  & 1408 & EPN \\
B1508+55 & 225 & GMRT & B1848+08 & 445 & AO &  & 1642 & EPN \\
& 408 & EPN & B1857$-$26 & 225 & GMRT &  &  &  \\
& 610 & EPN & B1859+03 & 1225 & AO &  &  &  \\
& 925 & EPN &  & 2250 & AO &  &  &  \\
& 1408 & EPN &  &  &  &  &  &  \\
& 1642 & EPN &  &  &  &  &  &  \\
&  &  &  &  &  &  &  &  \\ \hline
\end{tabular}
\end{center}
\end{table*}

Our analysis followed these steps: (1) Identification of core
components in profiles. We have used several established methods for
discriminating the core components: (i) quasi--centrality of the component
flanked by one or two pairs of conal components, (ii) associating the core
component with the inflexion point of the polarisation position angle
traverse and sense--reversing signature of circular polarisation under the
core component (although the polarimetry was not always available in our
data), (2) identification of conal pairs as leading and trailing components,
primarily based on symmetric location of components on either side of the
core one, (3) estimation of phase shifts according to eqn. (\ref{delta_phi}),
and (4) estimation of the emission heights according to eqn. (\ref{r_del}).
This simple procedure is not always straightforward in practice. There are cases when
components merge (especially as a function of the observing frequency) and
their positions are not clearly determined. In these cases we decided to use
a Gaussian fitting method introduced by Kramer et. al (1994). For
consistency, we applied this method to all our profiles, even those in which
all components were clearly distinguishable. The Gaussian fitting procedure
was as follows: fix Gaussian parameters of identifiable components
(position, amplitude, width), put additional Gaussian with some initial
parameters and allow the fitting method to find the best position,
amplitudes and widths. For detailed information about Gaussian fitting
method see Kramer et. al (1994). After applying this method to unresolved
components we obtained a full and consistent set of component positions to
be used in further analysis.

\section{Results}

\label{sec5} Our preliminary data set was reduced to 23 pulsars, because it
was hard to perform a reasonable fitting on 5 pulsars (B0950+08, B1859+03,
B1946+35, B2020+28, B2315+21) due to either low S/N or problems with
determining positions of core and/or conal components, even using a Gaussian
fitting method. The determined positions of all components in a pulse
profile, using the already mentioned Gaussian fitting procedure, were used
for a further analysis.

Firstly, we estimate the phase--shifts due to the A/R effect, and find the
emission height $r_\mathrm{AR}$ using eqn. (\ref{r_del}). For calculating the
phase--shifts using eqn. (\ref{delta_phi}), we first identify the leading
and trailing component of a cone of emission using the location
of the peak of the Gaussian fitted component as $\phi_l$ and $\phi_t$ (this
profile measure is often called as peak--to--peak component separation (PPCS),
see for e.g. G\&Ga and G\&Gb, Dyks et al. 2004, Mitra \& Li 2004). Correspondingly
we denote the phase shift as $\Delta\phi_{PPCS}$. Wherever possible,
$\Delta\phi_{PPCS}$ was estimated for the inner and outer cones, separately.

To estimate errors of single phase position measurements we used the formula
introduced by Kijak \& Gil (1997):
\begin{equation}  \label{sigma}
\sigma = \tau \times \sqrt{1+\left(\frac{\sigma_\mathrm{I}}{\mathrm{I}}\right)^2},
\end{equation}
where $\tau$ is the observing resolution, $\sigma_\mathrm{I}$ is the root mean square of total intensity, and I is the
measured signal level. Since $W$ estimations need two measurements (left and
right side of profile at particular intensity level) and $\Delta\phi$
estimations need three measurements ($\phi_0$, $\phi_\mathrm{l}$ and $\phi_%
\mathrm{t}$) we took
\begin{equation}  \label{sigma2}
\sigma_\mathrm{W} = \sqrt{\sigma_\mathrm{l}^2 + \sigma_\mathrm{r}^2}
\end{equation}
\begin{equation}  \label{sigma3}
\sigma_{\Delta\phi} = \sqrt{\sigma_{\phi_0}^2 + \sigma_{\phi_\mathrm{l}}^2 +
\sigma_{\phi_\mathrm{t}}^2}
\end{equation}
as an error for $W$ and $\Delta\phi$ respectively.

\begin{figure}
\begin{center}
\includegraphics[width=6.5cm]{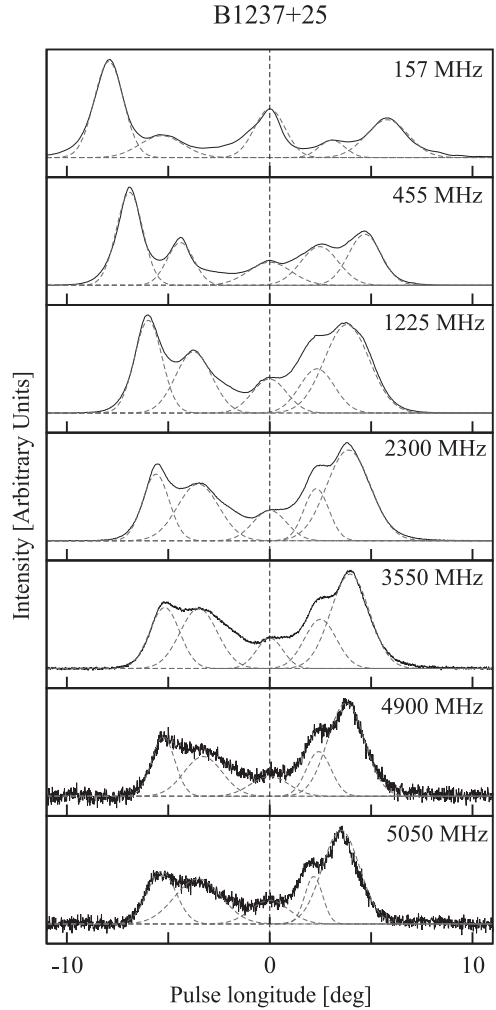}
\end{center}
\caption{ The profile evolution and fitted Gaussians of B1237+25 through a
number of frequencies. This is a good example showing clear A/R effects at all frequencies. }
\label{b1237}
\end{figure}

The best example of a pulsar whose apparent components are very well
reproduced by the Gaussian fitting method is PSR B1237+25 presented in Fig.
(\ref{b1237}) at a number of frequencies. All B1237+25 profiles are fitted
perfectly with 5 Gaussian components, being an ideal example of
applicability of the Gaussian fitting technique used in this pulsar. This
pulsar clearly shows the A/R effects at all frequencies,
consistently over the spectrum from about 5 GHz to about 0.16 GHz (see
also Table A.5.2 in Section A.5 in the on--line Appendix\footnote{http://astro.ia.uz.zgora.pl/$\sim$chriss/Krzeszowski\_2009\_appendix.ps}). The profiles are arranged
in panels with ascending frequency. The original profiles are plotted with
solid lines, while the fitted Gaussian components are drawn with dashed
lines. The profiles are aligned with respect to the peak of the central
Gaussian component identified as the core component. All the profiles of
pulsar B1237+25 show a clear A/R shift. Srostlik \& Rankin (2005) have found
that PA steepest gradient traverse lags the sign changing circular polarisation
zero point by about 0.4$^\circ$. However if one takes this value into account, B1237+25
still shows the A/R shift.

Our main results, which will be discussed later on, show the commonness of A/R effects in pulsar profiles.
Within our sample of 23 pulsars 7 show
clear A/R effects, 12 of them (doubtful cases) show a clear tendency
towards this effect, while the remaining 4 are so--called counter examples.
When it comes to single $\Delta \phi$ measurements we found that 45 measurements show
negative shifts, which is expected for A/R effects, 23 measurements are doubtful within
uncertainties, while remaining 19 show opposite shifts. However taking into account only the outer cones
(where more than one cone is seen) reveals that 30 shift measurements show A/R effects,
18 are doubtful cases whereas 14 show no A/R effects at all. For the majority of non--A/R cases, the problem lies
mostly in misidentifying core components. Some of the cases are partial cones which also makes
it difficult to measure A/R shift properly. For a detailed discussion of various issues regarding A/R
calculations see section (\ref{sec3}). For a discussion of the results see section (\ref{sec6}).

Note that all individual results are available in the
form of figures and detailed tables in the electronic form as an appendix to this paper.

\section{Conclusions and discussion}

\label{sec6} 
\begin{figure}
\begin{center}
\includegraphics[width=10cm,angle=270]{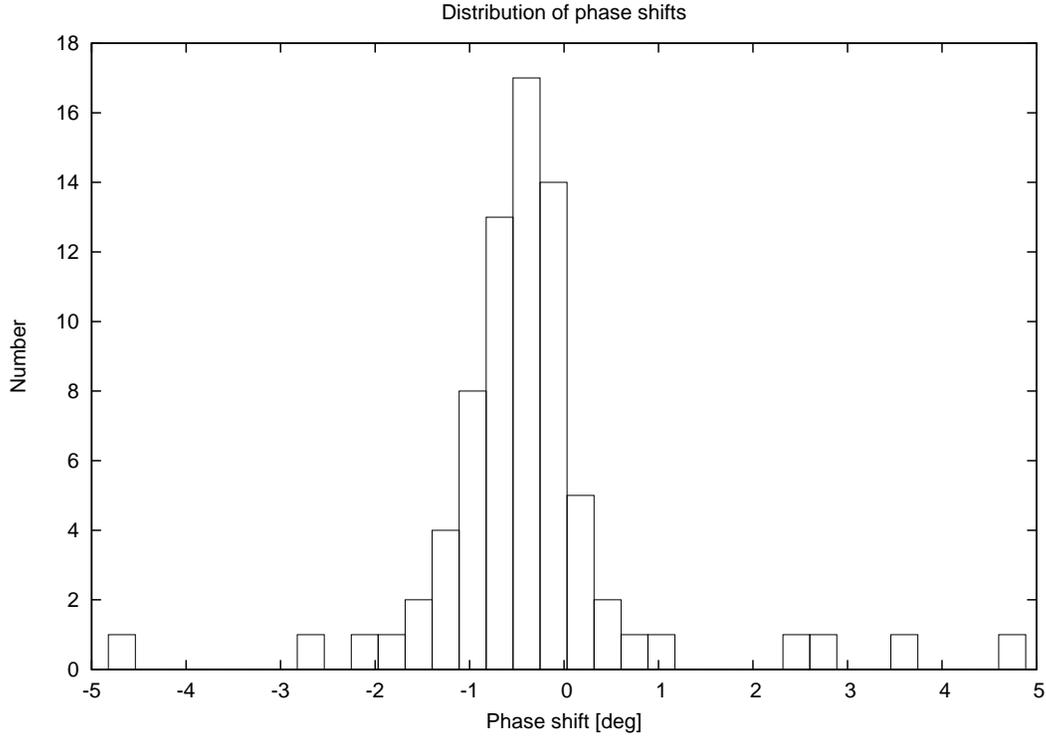}
\end{center}
\caption{ Distribution of measured phase shifts $\Delta\protect\phi_{PPCS}$. Note that a clear tendency for negative shifts is seen in the sample. }
\label{hist}
\end{figure}

Observationally, the peak of the core component lagging the
centre of the outer conal component has always been considered as an
evidence for A/R effects operating in pulsars. Those effects have been
demonstrated in a few pulsars so far: PSR B0329+54 by Malov \& Suleimanova (1998)
and G\&Ga between 103 MHz to 10.5 GHz and PSR's B0450$-$18, B1237+25,
B1821+05, B1857$-$26, B2045$-$16 and PSR B2111+46 at 325 MHz by G\&Gb. In this
work, our aim was to verify the A/R effects for a large sample of
pulsars. While our sample consists of 23 pulsars, we note that our analysis
is aided by using multi--frequency measurements, where each frequency serves
as an independent data set. Hence, although our sample consists of three
pulsars common with G\&Gb, the measurements are done at different
frequencies and can be treated as independent check for their results. In our
data set of 62 pulse profiles measured, the A/R effects are clearly seen in 30
profiles. Fig.~(\ref{hist}) shows the distribution of the measured phase shifts
$\Delta\phi_{PPCS}$ for our sample of pulsars (for the data see the Appendix).
 The tendency for the negative shift, which is the signature for A/R
effect, is clearly visible.

Out of the four counter--examples noted in our sample, the most
outstanding one is PSR B1831$-$04 (Fig. \ref{b1831}). We think that
the central component in this pulsar might be incorrectly identified
as the core emission (as discussed in section 3). The flat PA
traverse seems to suggest that the line of sight is not cutting the
overall beam centrally (see another example of broad profile pulsar
B0826$-$34 discussed by Gupta, Gil, Kijak et al. 2004), so the core
emission might be missing. If this is the case then the central
component could arise from grazing the third, innermost cone of
emission, although its non-central location is a concern in this model. 
However, if our hypothetical 3--rd cone interpretation is
incorrect, then B1831$-$04 remains the strongest counter--example of
A/R effects.

The three other counter--examples B1738$-$08, B1848+04 and B1859+03 show
a marginal deviation from the A/R effects and are probably also related to our
inability of identifying the location of the core component. In the first
case (B1738$-$08) the core component is not apparent, although a blind
Gaussian fitting method indicates a central ``core" component (see Fig. A.9 in the
on--line Appendix). The second pulsar (B1848+04) is again a typical example of a
broad profile pulsar (see Fig. A.13 in the on--line Appendix), where the
core component is likely to be missed. The third pulsar (B1859+03) (see Fig. A.4 in the on--line Appendix)
is a clear example of a
pulsars with missing component (in this case the trailing one).
Therefore, we conclude, that all 4 cases of strong counter--examples reported
are the result of mis--identification of the core component. Sometimes weak
conal emission or merged conal components leads to wrong
identification of the emission components. For example in PSR B1541+09 the
leading conal component at 1225 and 2330 MHz is difficult to model. In cases
like PSR B1944+17 and B1737+13 the weak conal emission poses difficulties in determining
the location of the components.
It is likely that the same set of problems occur in the group of doubtful cases (i.e. pulsars
which show A/R effects at some frequencies but not in others).
This strengthens our general conclusions that A/R effects are apparent in
pulsar profiles.

\begin{figure}
\begin{center}
\includegraphics[width=10cm]{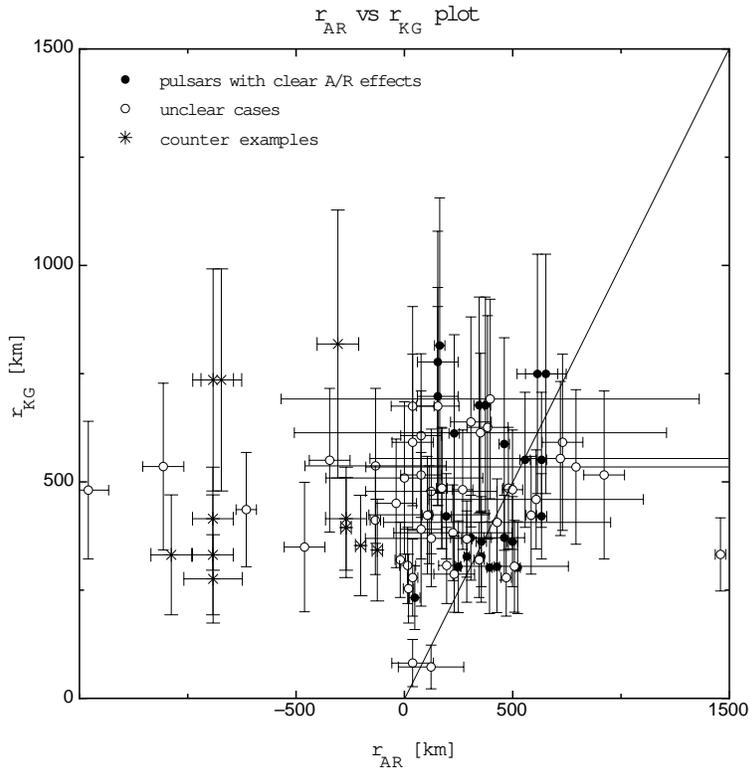}
\end{center}
\caption{ Relation between the delay emission height $r_\mathrm{AR}$ given
by eqn.~(\protect\ref{r_del}) and the semi--empirical geometrical height
estimate $r_\mathrm{KG}$ given by eqn.~(\protect\ref{rkg}). Note that
statistically $r_\mathrm{KG} > r_\mathrm{AR}$ as seen by the excess of points
lying above the $r_\mathrm{AR}~=~r_\mathrm{KG}$ line in the figure (see legend to distinguish different
groups of pulsars). See text for details. }
\label{r_em_plot}
\end{figure}

In cases where A/R effects are seen between the core and conal components
in pulsars, implies that the core emission necessarily has to originate
closer to the stellar surface than the conal emission (see for e.g. G\&Ga,
Dyks et al. 2004). The delay emission
heights obtained by eqn.~(\ref{r_del}) are essentially the conal emission
heights with respect to the core emission. Alternatively geometrical methods
can be used to find the conal emission height $r_\mathrm{KG}$ as specified by eqn.~%
(\ref{rkg}). A comparison between the conal emission heights measured by these
two methods is shown in Fig.~(\ref{r_em_plot}). This plot can be compared to Fig. (29) of BCW where a similar effect
is noted, although the delay heights there were measured from polarisation
observations and they used log--log axes. In this figure the clear A/R cases
are marked with filled circles, doubtful cases are marked with open circles and
the so called counter--examples are marked with stars. Note that many of the negative $r_\mathrm{AR}$ values, in Fig.~(\ref{r_em_plot}) correspond to the counter--examples in our sample. 
It is noteworthy that
statistically the relation $r_\mathrm{KG} > r_\mathrm{AR}$ seems to
hold well. 
 There are primarily two reasons which can explain this effect.
Firstly the $r_\mathrm{AR}$ values can be underestimated if the putative
core emission arises from a definite height above the polar cap. In order to
square the two height estimates by making the cluster of points in
Fig.~(\ref{r_em_plot}) to be located on the 45$^{\circ }$ line, a nominal average core
emission height of about 100 km is needed. Alternatively, the geometrical
emission height $r_\mathrm{KG}$ might be overestimated, since the
assumption prevailing in estimating these heights is that the measured
widths entail the last open field lines, i.e. $s=1$ (see eqn.~\ref{rho}).
If the parameter $s$ was around 0.7, the two height estimates would square
with each other. As discussed by Mitra \& Li (2004), a careful and systematic
study of the relation of the core emission w.r.t. the polarisation position
angle traverse might help to resolve this issue.

\section*{Acknowledgement}

We thank the GMRT operational staff for observational support.
The GMRT is run by the National Centre for Radio Astrophysics of
the Tata Institute of Fundamental Research.
We acknowledge the usage of the Gaussian fitting program given to us by Michael
Kramer which has been extensively used in this paper. JG and JK acknowledge
partial support of the Polish Grant N N203 2738 33. JK and KK acknowledge
the Polish Grant N 203 021 32/2993. This paper was also partially supported by
the Polish Grant N N203 3919 34. We would like to thank an anonymous referee whose
comments and suggestions helped us to improve our paper. We thank Urszula Maciejewska
for critical reading of the manuscript.

\label{lastpage}
\end{document}